\documentclass[%
aip,
amsmath,amssymb,
reprint,%
]{revtex4-1}
\usepackage{graphicx}
\usepackage{dcolumn}
\usepackage{mathrsfs}
\usepackage{bm}
\usepackage{lineno}
\usepackage{hyperref}
\usepackage{bbm}
\usepackage{ulem}
\usepackage{color}

\newcommand{\cX}{{\cal X}}
\newcommand{\cY}{{\cal Y}}
\begin{document}

\title{Quantum dense metrology by an SU(2)-in-SU(1,1) nested interferometer}
\author{Wei Du}
 \email{weiduecnu@126.com}
\affiliation{Quantum Institute of Light and Atoms,Department of Physics, East China Normal University, Shanghai, 200241, People's Republic of China}
\author{J. F. Chen}
\affiliation{Shenzhen Institute for Quantum Science and Engineering and Department of Physics, Southern University of Science and Technology, Shenzhen 518055, China}
\author{Z. Y. Ou}
 \email{zou@iupui.edu}
\affiliation{Department of Physics, Indiana University-Purdue University Indianapolis, Indianapolis, IN 46202, USA}
\author{Weiping Zhang}
 \email{wpz@sjtu.edu.cn}
\affiliation{Department of Physics and Astronomy, Tsung-Dao Lee Institute, Shanghai Jiao Tong University, Shanghai 200240, People's Republic of China}

\begin{abstract}
With the help of quantum entanglement, quantum dense metrology (QDM) is a technique that can perform the joint estimates of two conjugate quantities such as phase and amplitude modulations of an optical field with an accuracy beating the standard quantum limit simultaneously. SU(1,1) interferometers (SUI) can realize QDM with detection loss tolerance but is limited in absolute sensitivity. Here we present a QDM scheme with a linear interferometer (SU(2)) nested inside an SU(1,1) interferometer. By using a degenerate SUI and controlling the phase angle of the phase-sensitive amplifiers in SUI, we can achieve the optimum quantum enhancement in the measurement precision of arbitrary mixture of phase and amplitude modulation.

\end{abstract}

\maketitle

It is well-known \cite{Caves} that the phase uncertainty of an interferometer with classical sources is bounded by standard quantum limit (SQL) that scales as $ 1/\sqrt{N} $ with $ N $ as the photon number sensing the phase, which is viewed as classical resource. That is why strong power of light is employed inside the Laser Interferometer Gravitational-Wave Observatory (LIGO) \cite{Abbott1} for the highest absolute sensitivity. But high power causes various problems such as thermal effects. In the field of biological imaging, light toxicity of the probe field is a huge issue \cite{Bowen}.
A possible way to reach a high absolute sensitivity of phase measurement but at a relatively low light level is to reduce the vacuum fluctuation by placing the unused input port of the interferometer in a squeezed state of light \cite{Caves,Xiao,gra,Abadie,Yonezawa,Schnabel}.

A quantity that is equally important at fundamental level but relatively less in practical applications is the amplitude of a field, whose noise can also be reduced by the squeezed states \cite{xiao2}. An example is the radiation pressure noise in LIGO induced by light intensity fluctuations \cite{Caves}. However, because of the Heisenberg uncertainty principle, it is impossible to beat SQL with a single-mode squeezed state in the measurement of phase and amplitude simultaneously.
In recent years, with the application of quantum entanglement, two-mode squeezed light is demonstrated to be capable of embedding two or more non-commuting observables in information with their measurement precision beyond the SQL simultaneously. This is so-called quantum dense coding  in quantum information science \cite{braun00} or quantum dense metrology (QDM) in quantum metrology \cite{zhang,Li,Steinlechner,Jiamin,Yuhong}.

An alternative way to achieve quantum-enhanced sensitivity is to amplify noiselessly the signal\cite{Ou,Corzo,Kong}, achieved in the so-called SU(1,1) interferometer (SUI) \cite{pl,ou12,jing,F,Mathieu,Brian}, which is firstly proposed by Yurke {\it et al.}\cite{Yurke} some thirty years ago. SUIs, while achieving a quantum-enhanced phase measurement, possess a number of advantages \cite{ou-li} to the conventional squeezed state interferometers especially the property of detection loss tolerance \cite{ou12,Marino,Mathieu,li19}. Recently, it was demonstrated that SUI is also capable of measuring multiple noncommuting parameters and beating the standard quantum limit simultaneously \cite{Jiamin,Yuhong}. In spite of so many advantages over the conventional interferometers, SUIs suffer low phase sensing photon numbers ($ I_{ps} $) due to saturation of parametric amplifiers. Thus the absolute sensitivity of such type of interferometer is still not comparable to classical interferometers. Moreover, current schemes of SUI  have so far only the phase measurement that makes full use of available resource and reaches the optimum sensitivity but not amplitude \cite{Jiamin,liu19,assad}.

In this paper, we consider a more practical scheme, embedding a linear interferometer (also known as SU(2) interferometer \cite{Yurke}) inside an SU(1,1) interferometer. Although the linear interferometer has strong injection for increasing absolute sensitivity, it operates in dark fringe mode so that the SUI basically works with no injection and thus low photon numbers inside.  We will demonstrate that the scheme is suitable for simultaneous measurement of phase and amplitude, thus realizing QDM,  but without the limitation on photon number, and its degenerate version can be used to measure either phase or amplitude modulation signal, or any arbitrary mixture of the two,  with an optimum sensitivity that is reached with full use of both classical and quantum resources.

Before discussing QDM, let us first define the phase (PM) and amplitude (AM) measurement.
A phase modulation signal $\delta$ can be added to an incoming probe field $\hat a_{in}$ by a phase factor: $e^{i\delta}\hat a_{in}$ without adding extra vacuum noise. But since amplitude modulation $\epsilon$ changes the intensity or energy of the probe field, which is equivalent to a loss, we need to model it with a beam spitter:
\begin{eqnarray}\label{amloss}
\hat{a}=t \hat{a}_{in}+r\hat{a}_{\nu}
\end{eqnarray}
with $ t= e^{-\epsilon}$ and $ r=\sqrt{1-t^2}$. Here, $\hat{a}_{\nu}$ is the vacuum field coupled in through loss. But if the modulation signal $\epsilon$ is very small: $\epsilon \ll 1$ so that $t\approx 1-\epsilon, r\approx \sqrt{2\epsilon}$, we can neglect the vacuum contribution because its noise $\langle \Delta^2X\rangle = t^2 \langle \Delta^2X_{in}\rangle + r^2 \langle \Delta^2X_{\nu}\rangle\approx (1-2\epsilon)\langle \Delta^2X_{in}\rangle + 2\epsilon \langle \Delta^2X_{\nu}\rangle \approx \langle \Delta^2X_{in}\rangle$. With this in mind, we will drop the vacuum term in Eq.(\ref{amloss}) in all treatment of amplitude modulation later. Therefore, for small $\delta,\epsilon (\ll1$), we have the modulated field as $ \hat{a}'\approx e^{i\delta}e^{-\epsilon}\hat{a}_{in}\approx(1+i\delta-\epsilon)\hat{a}_{in} $

\begin{figure}[t]
	\includegraphics[width=8.5cm]{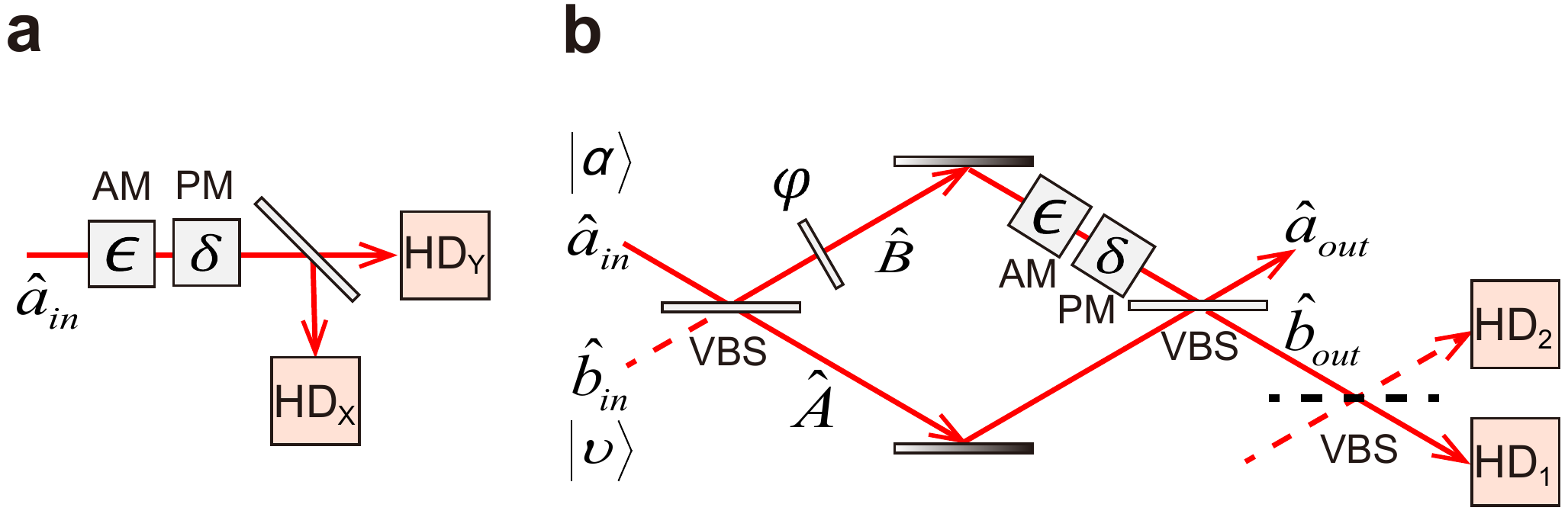}
	\caption{Schematic diagram for (a) direct homodyne measurement;(b) interferometer with homodyne measurement. VBS: variable beam splitter; HD: homodyne detection; AM: amplitude modulation; PM: phase modulation.}
	\label{fig:SQI}
\end{figure}

We start by first considering how to make simultaneous AM and PM measurement with a classical coherent state.  The simplest scheme is by direct homodyne measurement, as shown in Fig.\ref{fig:SQI}(a) and simultaneous measurement is achieved by splitting the modulated field with a beam splitter. An alternative way is by a linear or SU(2) interferometer, as shown in Fig.\ref{fig:SQI}(b) with similar signal splitting.  Both schemes have been shown \cite{li19} to achieve optimum  sensitivity in joint measurement by a classical field. But the interferometric scheme is advantageous to the direct homodyne detection scheme because the former can operate at the dark port without a large coherent component and it is especially suitable for increasing the intensity of the probe field for higher sensitivity. To prepare the discussion on QDM with SU(1,1) interferometers, we analyze in detail the interferometric scheme next.

For the interferometric scheme in Fig.\ref{fig:SQI}(b), we can find the output field by using the following beam splitter relations (refer to Fig.\ref{fig:SQI}(b) for notations):
\begin{eqnarray}\label{SQ-MZI} \nonumber
\hat{A}&=&\sqrt{T_{1}}\hat{a}_{in}+\sqrt{R_{1}}\hat{b}_{in},~~ \hat{B}=\sqrt{T_{1}}\hat{b}_{in}-\sqrt{R_{1}}\hat{a}_{in},\nonumber\\
&&\hskip 0.1in \hat{a}_{out} = \sqrt{T_{2}}\hat{A}-\sqrt{R_{2}}\hat{B}e^{-\epsilon}e^{i\varphi},\nonumber\\
&&\hskip 0.1in \hat{b}_{out} = \sqrt{T_{2}}\hat{B}e^{-\epsilon}e^{i\varphi}+\sqrt{R_{2}}\hat{A}.
\end{eqnarray}
The outputs are related to the input by
\begin{eqnarray} \nonumber
\label{eq:inoutMZI}
\hat{a}_{out}&=&(\sqrt{T_{1}T_{2}}+\sqrt{R_{1}R_{2}}e^{-\epsilon}e^{i(\varphi+\delta)})\hat{a}_{in}\nonumber\\& &~~~~+(\sqrt{R_{1}T_{2}}-\sqrt{T_{1}R_{2}}e^{-\epsilon}e^{i(\varphi+\delta)})\hat{b}_{in}\nonumber\\
\hat{b}_{out}&=&(\sqrt{T_{1}R_{2}}-\sqrt{R_{1}T_{2}}e^{-\epsilon}e^{i(\varphi+\delta)})\hat{a}_{in}\nonumber\\& &~~~~+(\sqrt{R_{1}R_{2}}+\sqrt{T_{1}T_{2}}e^{-\epsilon}e^{i(\varphi+\delta)})\hat{b}_{in},
\end{eqnarray}
where $\varphi$ is the phase added to field $B$ to account for overall phase difference in the interferometer.

For simplicity without loss of generality, we assume identical beam splitters: $T_1=T_2\equiv T,R_1=R_2\equiv R$. Notice that when $\varphi=0$,  the interferometer without modulations ($\delta=0=\epsilon$) acts as if nothing is there: $\hat a_{out} = \hat a_{in}$ and $\hat b_{out} = \hat b_{in}$. This is so-called dark fringe operation. We will work at this point throughout the paper. Now let input field $\hat a_{in}$ be in a coherent state $|\alpha\rangle$ and field $\hat b_{in}$ be in vacuum. Using small modulations assumption ($\delta,\epsilon \ll 1$),  we obtain for the dark port output:
\begin{eqnarray}\label{in-out-SU2}
\hat b_{out} \approx \hat b_{in} + \hat a_{in}\sqrt{TR} (\epsilon-i\delta).
\end{eqnarray}
Since coherent state has the same size of noise as vacuum state and $\delta,\epsilon \ll 1$, field $\hat b_{out}$ has fluctuations dominated by that of $\hat b_{in}$ but also contains information about modulations $\delta,\epsilon$.

Without loss of generality, we assume $\alpha =$ real for the input coherent state $|\alpha\rangle$ at $\hat a_{in}$. Then homodyne measurement of quadrature phase amplitudes $\hat X_{b_{out}} \equiv \hat b_{out}+ \hat b_{out}^{\dag}, \hat Y_{b_{out}} \equiv (\hat b_{out}- \hat b_{out}^{\dag})/i$ gives signal and noise respectively as
\begin{eqnarray}\label{SN-SU2}
\langle \hat X_{b_{out}} \rangle = 2 \alpha \epsilon\sqrt{TR},~&& \langle \hat Y_{b_{out}} \rangle = 2 \alpha \delta\sqrt{TR},\cr
\langle \Delta^2\hat X_{b_{out}} \rangle =~& 1&~= \langle \Delta^2 \hat Y_{b_{out}} \rangle ,
\end{eqnarray}
leading to the signal-to-noise ratio (SNR) for phase and amplitude modulations
\begin{eqnarray}\label{SNR-SU2}
SNR_{\delta} &\equiv &\langle \hat Y_{b_{out}} \rangle^2/\langle \Delta^2 \hat Y_{b_{out}} \rangle = 4TI_{ps}  \delta^2\cr
SNR_{\epsilon} &\equiv &\langle \hat X_{b_{out}} \rangle^2/\langle \Delta^2 \hat X_{b_{out}} \rangle = 4TI_{ps}  \epsilon^2,
\end{eqnarray}
where $I_{ps}\equiv R\alpha^2$ is the photon number for the probe field to the modulations, which can be considered as the overall classical resource for the measurement.
Optimum SNRs \cite{ou-li} of $4I_{ps}  \delta^2, 4I_{ps} \epsilon^2$ are achieved for a very unbalanced interferometer with $T\approx 1, R=1-T\ll 1$.

The above is for individual $\delta$ or $\epsilon$ measurement alone. For simultaneous measurement of $\delta$ and $\epsilon$, an extra BS ($T_3,R_3$) can be used to split the output but with split SNR as well: $SNR_{\delta}^{(s)}=4T_3I_{ps}  \delta^2, SNR_{\epsilon}^{(s)}=4R_3I_{ps} \epsilon^2$, similar to Fig.\ref{fig:SQI}(a). Notice that simultaneous measurement of $\delta$ and $\epsilon$ needs to share the overall resource of $I_{ps}$ ($T_3I_{ps}+R_3I_{ps} = I_{ps}$) \cite{li19}.

When squeezed states are used to replace the vacuum state at unused input port of $\hat b_{in}$,  the vacuum quantum noise can be reduced and the SNR in either phase \cite{Caves,Xiao,gra} or amplitude measurement\cite{xiao2} is enhanced, but not in both because only one quadrature is squeezed at a time. But with the employment of quantum entanglement, the technique of quantum dense metrology (QDM) is capable of measuring two non-orthogonal quadratures with quantum enhancement simultaneously\cite{Steinlechner}. SU(1,1) interferometers, which possess the property of detection loss tolerance,  also have the ability to measure simultaneously multiple quadrature-phase amplitudes at arbitrary angles with quantum enhanced precision\cite{Jiamin,Yuhong}. However, despite of its numerous advantages over traditional interferometers, SU(1,1) interferometers have limited absolute sensitivity due to practical issues. Here, to circumvent these problems, we study a variation of SU(1,1) interferometer, which combines a traditional linear or SU(2) interferometer  with an SU(1,1) interferometer for QDM and inherits the advantages of both interferometers.

\begin{figure}[t]
	\includegraphics[width=8cm]{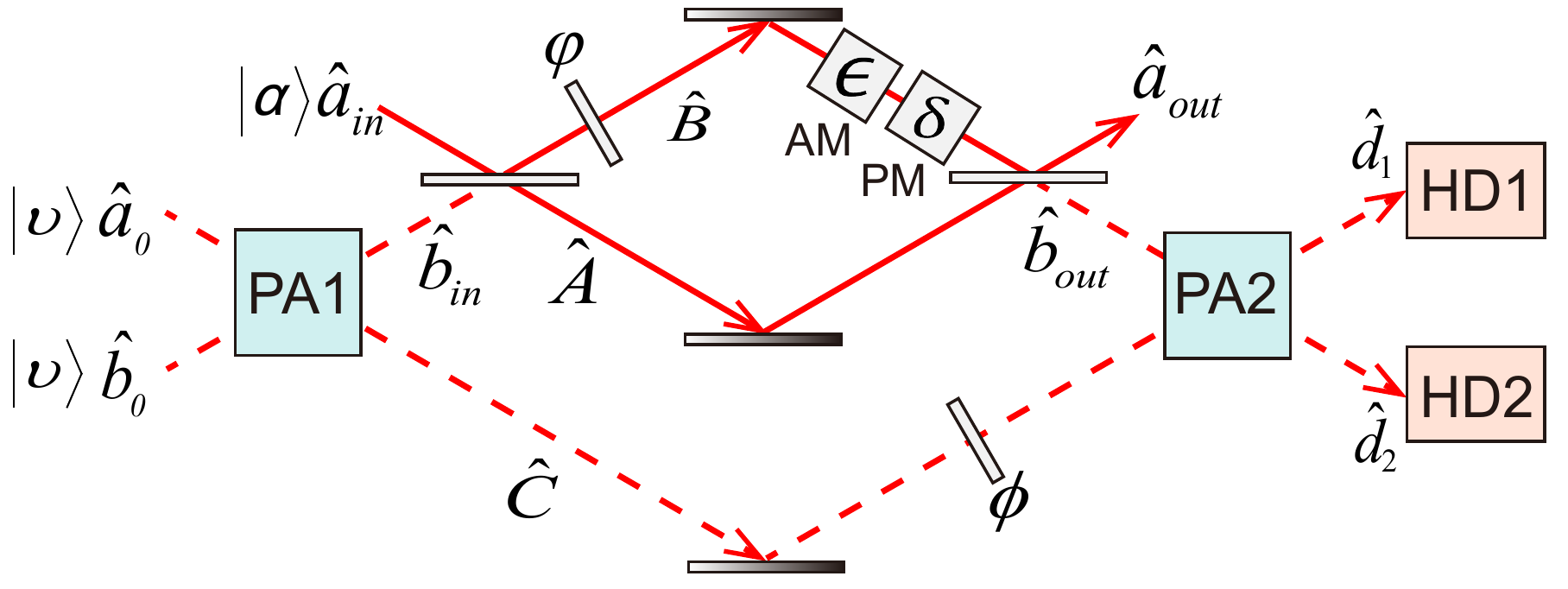}
	\caption{QDM with non-degenerate SU(1,1) interferometry. PA1, PA2: parametric amplifier; HD: homodyne detection.}
	\label{fig:SUIlinear}
\end{figure}

Consider the scheme shown in Fig.~\ref{fig:SUIlinear}. It is a Mach-Zehnder interferometer (MZI) nested in an SU(1,1) interferometer (SUI), which was shown recently \cite{du20} to possess all the advantages of SUI for phase measurement  but without the limit on the intensity of the probe field. To achieve QDM, we make homodyne measurement on both outputs of the second parametric amplifier (PA2), one for phase ($d_1$) and the other for amplitude ($d_2$). If the MZI works at dark fringe point, we have the input-output relation in Eq.(\ref{in-out-SU2}). Since $\hat b_{out} \approx \hat b_{in}$ as if MZI were not there, the overall performance of the SUI is not affected by MZI and it does not have any injection at its inputs ($\hat a_0,\hat b_0$) but the modulation signals $\delta,\epsilon$ is contained in $\hat b_{out}$ for SUI to measure.

Using the input-output relation for PA1 and PA2:
\begin{eqnarray}
\hat{b}_{in}=G_1\hat{b}_{0}+g_1\hat{a}_{0}^{\dagger},&&~ \hat{C}=G_1\hat{a}_{0}+g_1\hat{b}_{0}^{\dagger},\cr
\hat{d}_{1}=G_2\hat Ce^{i\phi} + g_2 \hat{b}_{out}^{\dag},&&~ \hat{d}_{2}=G_2 \hat{b}_{out}+ g_2\hat C^{\dag}e^{-i\phi},~~~~
\end{eqnarray}
we obtain the outputs of the SUI for QDM as
\begin{eqnarray}\nonumber
\hat{d}_{1}&=&(G_1G_{2}e^{i\phi}+g_1g_2)\hat{a}_{0}
\nonumber\\& &\hskip 0.1in+(g_1G_{2}e^{-i\phi}+G_1g_2)\hat{b}_{0}^{\dag} +g_2\sqrt{R}(\epsilon+i\delta)\hat{a}_{in}^{\dagger}\nonumber\\
\hat{d}_{2}\!&=&\!(G_1G_2+g_1g_{2}e^{-i\phi})\hat{b}_{0}
\nonumber\\& &\hskip 0.1in +(g_{1}G_{2}\!+\!G_{1}g_{2}e^{-i\phi})\hat{a}_{0}^{\dagger} +G_2\sqrt{R}(\epsilon-i\delta)\hat{a}_{in},~~~~
\end{eqnarray}
where $\phi$ is a phase to account for the overall phase of the SUI. Here we used Eq.(\ref{in-out-SU2}) with $T\approx 1$ for optimum performance of the MZI.

With input fields $ \hat{a}_{0} $ and  $ \hat{b}_{0} $ in vacuum state, $R\rightarrow 0$ and small modulations ($\delta,\epsilon \ll1$), it is straightforward to calculate the noise of the outputs of the SUI as
\begin{eqnarray}
&&\langle \Delta^2 \hat Y_{{d}_{1}}\rangle = \langle \Delta ^2\hat X_{{d}_{2}} \rangle \cr && \hskip 0.3in = |G_1G_2+g_1g_{2}e^{-i\phi}|^2
 +|g_{1}G_{2}\!+\!G_{1}g_{2}e^{-i\phi}|^2
 \cr && \hskip 0.3in = (G_1^2+g_1^2)(G_2^2+g_2^2) + 4G_1G_2g_1g_2 \cos\phi,
\end{eqnarray}
where $\hat Y_{{d}_{1}} = ({d}_{1}-{d}_{1}^{\dag})/i,\hat X_{{d}_{2}} = {d}_{2}+{d}_{2}^{\dag}$. The noise is minimum when $\phi=\pi$ corresponding to dark fringe. The signals are calculated as
\begin{eqnarray}
 \langle \hat Y_{{d}_{1}} \rangle =2g_2\delta \sqrt{I_{ps}}, \langle  \hat X_{{d}_{2}}\rangle = 2G_2\epsilon \sqrt{I_{ps}}
\end{eqnarray}
with $I_{ps}=R\alpha^2$. So, the SNRs for simultaneous measurement of $\delta$ and $\epsilon$ are
\begin{eqnarray}\nonumber
&&SNR_{SUI}(\hat{Y}_{{d}_{1}})=\frac{4g_{2}^{2}I_{ps}\delta^{2}}
{(G_{2}G_{1}-g_{1}g_{2})^{2}+(G_{1}g_{2}-G_{2}g_{1})^{2}}\\
&&SNR_{SUI}(\hat{X}_{{d}_{2}})=\frac{4G_{2}^{2}I_{ps}\epsilon^{2}}
{(G_{2}G_{1}-g_{1}g_{2})^{2}+(G_{1}g_{2}-G_{2}g_{1})^{2}}.\nonumber
\label{eq:SUISNR}
\end{eqnarray}
Both reach optimum value when $G_2\approx g_2\gg 1$:
\begin{eqnarray}
&&SNR_{SUI}^{(op)}(\hat{Y}_{{d}_{1}})=2I_{ps}\delta^{2}(G_1+g_1)^2\nonumber\\
&&SNR_{SUI}^{(op)}(\hat{X}_{{d}_{2}})=2I_{ps}\epsilon^{2}(G_1+g_1)^2.
\label{eq:SUISNRop}
\end{eqnarray}
When $(G_1+g_1)^2/2 > 1$, both measurement of $\delta$ and $\epsilon$ beats simultaneously the classical sensitivity expressed in Eq.(\ref{SNR-SU2}) by the same factor of $(G_1+g_1)^2/2$, thus achieving QDM. Notice that the total resource of $I_{ps}(G_1+g_1)^2$ is shared equally, satisfying the quantum resource sharing law\cite{li19,assad}. The resource now consists of photon number $I_{ps}$ and quantum entanglement characterized by $(G_1+g_1)^2$.

\begin{figure}[t]
	\includegraphics[width=8.5cm]{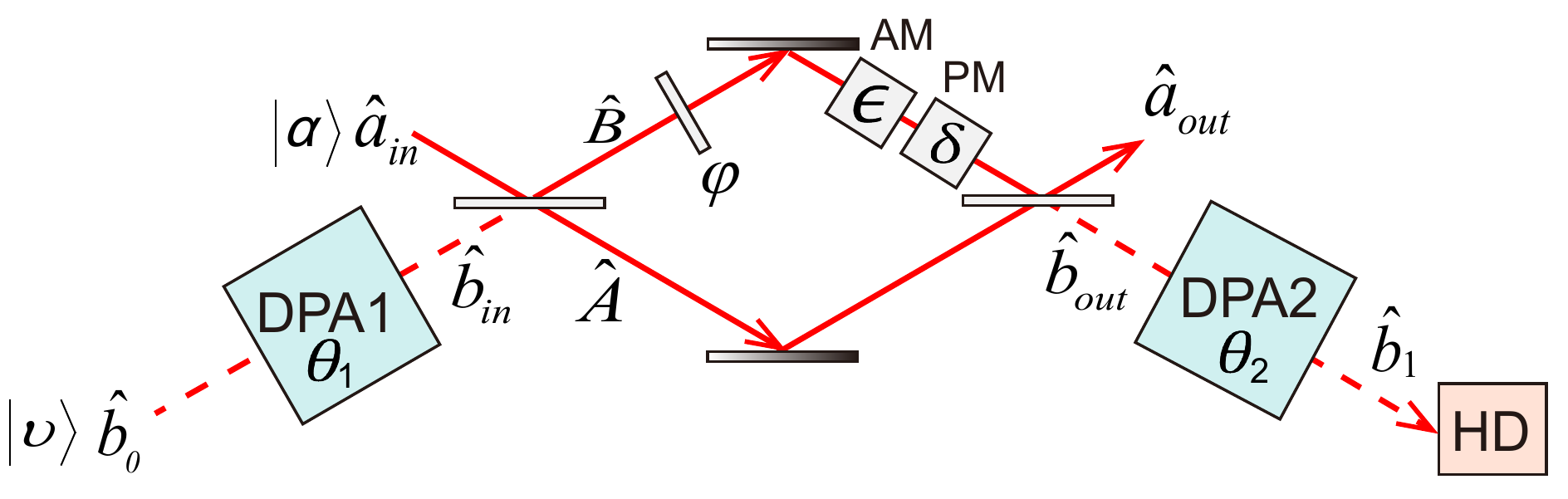}
	\caption{QDM with degenerate SU(1,1) interferometry when loss is considered. DPA: Degenerate parametric amplifier.}
	\label{fig:DPA}
\end{figure}

Since the resource is shared equally between phase and amplitude measurement, when SUIs are used for phase or amplitude measurement alone, only half the optimum sensitivity is realized. Recently, a variation of SUI \cite{liu19} employs dual beam to sense the modulations for taking all the resource for phase measurement and achieving optimum sensitivity. However, the scheme is only suitable for phase.
In practice, the information embedded may not be purely in phase but could be in amplitude or something of a mixture of phase and amplitude. To achieve optimum sensitivity for these, we make a variation to the SUI. We use a degenerate parametric amplifier (DPA) instead of the regular non-degenerate PA, as shown in Fig.\ref{fig:DPA}. Benefiting from the degenerate PA, which can be regarded as a phase-sensitive amplifier, we can selectively distribute all resources to any mixture of phase and amplitude for optimum measurement sensitivity, as shown next.

Referring to the notations in Fig.\ref{fig:DPA}, we write the input-output relations of DPA as
\begin{eqnarray}\label{eq:DOPA12}
&&\hat{b}_{in}=G_{1}\hat{b}_{0}+g_{1}e^{i\theta_1}\hat{b}_{0}^{\dagger},\cr
&&\hat{b}_{1}=G_{2}\hat{b}_{out}+g_{2}e^{i\theta_2}\hat{b}_{out}^{\dagger}
\end{eqnarray}
where phase $\theta_{1(2)}$ is for DPA1(2). Using Eq.(\ref{in-out-SU2}) with $T\approx 1$ for $\hat b_{out}$, we find the output of DPA2 as
\begin{eqnarray}\label{eq:DPA12}
\hat{b}_{1}&=& G_{2}[G_{1}\hat{b}_{0}+g_{1}e^{i\theta_1}\hat{b}_{0}^{\dagger}+\hat a_{in}\sqrt{R}(\epsilon-i\delta)]\cr
&&\hskip 0.1in +g_{2}e^{i\theta_2}[G_{1}\hat{b}_{0}^{\dagger}+g_{1}e^{-i\theta_1}\hat{b}_{0}+\hat a_{in}^{\dagger}\sqrt{R}(\epsilon+i\delta)]\cr
&=& G_T \hat{b}_{0}+ g_T \hat{b}_{0}^{\dagger} + \sqrt{R} G_2 (\epsilon-i\delta)\hat a_{in}
\cr && \hskip 0.4in+ \sqrt{R} g_{2}e^{i\theta_2}(\epsilon+i\delta)\hat a_{in}^{\dagger}],~~~~
\end{eqnarray}
where $G_T \equiv G_{1}G_{2} + g_{1}g_{2}e^{i(\theta_2-\theta_1)}, g_T \equiv g_{1}G_{2}e^{i\theta_1} + G_{1}g_{2}$ $e^{i\theta_2}$ as the overall gains of the degenerate SUI.
Since $\hat b_0$ is in vacuum, the signal part of the output is then
\begin{eqnarray}\label{eq:DPA1}
\langle \hat{b}_{1}\rangle &=& \alpha \sqrt{R} [G_2 (\epsilon-i\delta)+ g_{2}e^{i\theta_2}(\epsilon+i\delta)]\cr
&=& \alpha \sqrt{R} e^{i\theta_2/2} \big[\epsilon (g_{2}e^{i\theta_2/2}+G_2 e^{-i\theta_2/2})\cr &&  \hskip 0.5in + i \delta (g_{2}e^{i\theta_2/2}-G_2 e^{-i\theta_2/2})\big].
\end{eqnarray}
If we measure $\hat \cX\equiv \hat b_1e^{-i\theta_2/2}+\hat b_1^{\dag} e^{i\theta_2/2}, \hat \cY\equiv (\hat b_1e^{-i\theta_2/2}-\hat b_1^{\dag} e^{i\theta_2/2})/i$, we find the signal becomes
\begin{eqnarray}\label{eq:DPA2}
\langle \hat \cX\rangle
&= & -2 (G_2+g_2)\sqrt{I_{ps}} \gamma_- \cr
\langle \hat \cY\rangle
&= & -2(G_2-g_2)\sqrt{I_{ps}}\gamma_+,
\end{eqnarray}
where $\gamma_{-} \equiv -\epsilon \cos\frac{\theta_2}{2}+\delta \sin \frac{\theta_2}{2}, \gamma_+\equiv \epsilon \sin\frac{\theta_2}{2}+\delta \cos \frac{\theta_2}{2}$ are two orthogonal modulation signals.
Therefore, measurement of $\hat \cX, \hat \cY$ gives orthogonal quantities $\gamma_{\mp}$ with $\gamma_{-}$ amplified by $G_2+g_2$ but $\gamma_{+}$ de-amplified by $G_2-g_2$.

For the noise of the degenerate SUI,
since $R\ll 1$, it  is governed by $G_T,g_T$ and with $\hat b_0$ in vacuum, we have
\begin{eqnarray}
\langle \Delta ^2\hat \cX \rangle &=& |G_Te^{-i\theta_2/2}+g_T^*e^{i\theta_2/2}|^2\cr
&=& (G_2+g_2)^2|G_1+g_1e^{-i\Delta}|^2\cr
\langle \Delta^2 \hat \cY\rangle &=& |G_Te^{i\theta_2/2}-g_T^*e^{i\theta_2/2}|^2 \cr
&=& (G_2-g_2)^2|G_1-g_1e^{-i\Delta}|^2,
\end{eqnarray}
where $\Delta \equiv \theta_1-\theta_2$ is total phase of the SUI.
With $\Delta = \pi$ for dark fringe (destructive interference), we have
\begin{eqnarray}\label{N-DSUI}
\langle \Delta ^2\hat \cX \rangle &=& (G_2+g_2)^2(G_1-g_1)^2\cr
 \langle \Delta^2 \hat \cY\rangle &=& (G_2-g_2)^2(G_1+g_1)^2.
\end{eqnarray}
So, the SNRs for the measurement of $\hat \cX, \hat \cY$ are
\begin{eqnarray}
SNR_{\cX} &=& 4I_{ps}\gamma_-^2 (G_1+g_1)^2 \cr
 SNR_{\cY} &=& 4I_{ps}\gamma_+^2 (G_1-g_1)^2.
\end{eqnarray}
Because there is only one output for the degenerate SUI, we cannot make simultaneous measurement of the orthogonal quantities $\gamma_{\pm}$ and so the degenerate SUI is not suitable for QDM. But a homodyne detection of $\cX$ at the output of the degenerate SUI will give the measurement of $\gamma_-=-\epsilon\cos\theta_2/2 + \delta\sin \theta_2/2$, an arbitrary mixture of AM ($\epsilon$) and PM ($\delta$) signals, depending on $\theta_2$ with a quantum enhancement of $(G_1+g_1)^2$. Setting $\theta_2=0$ gives the full amplitude modulation ($\epsilon$) whereas $\theta_2=\pi$ gives the full phase modulation ($\delta$). The measurement reaches the optimum measurement sensitivity making full use of the quantum resource of $I_{ps}(G_1+g_1)^2$ of the probe field.

\begin{figure}[t]
	\includegraphics[width=8.5cm]{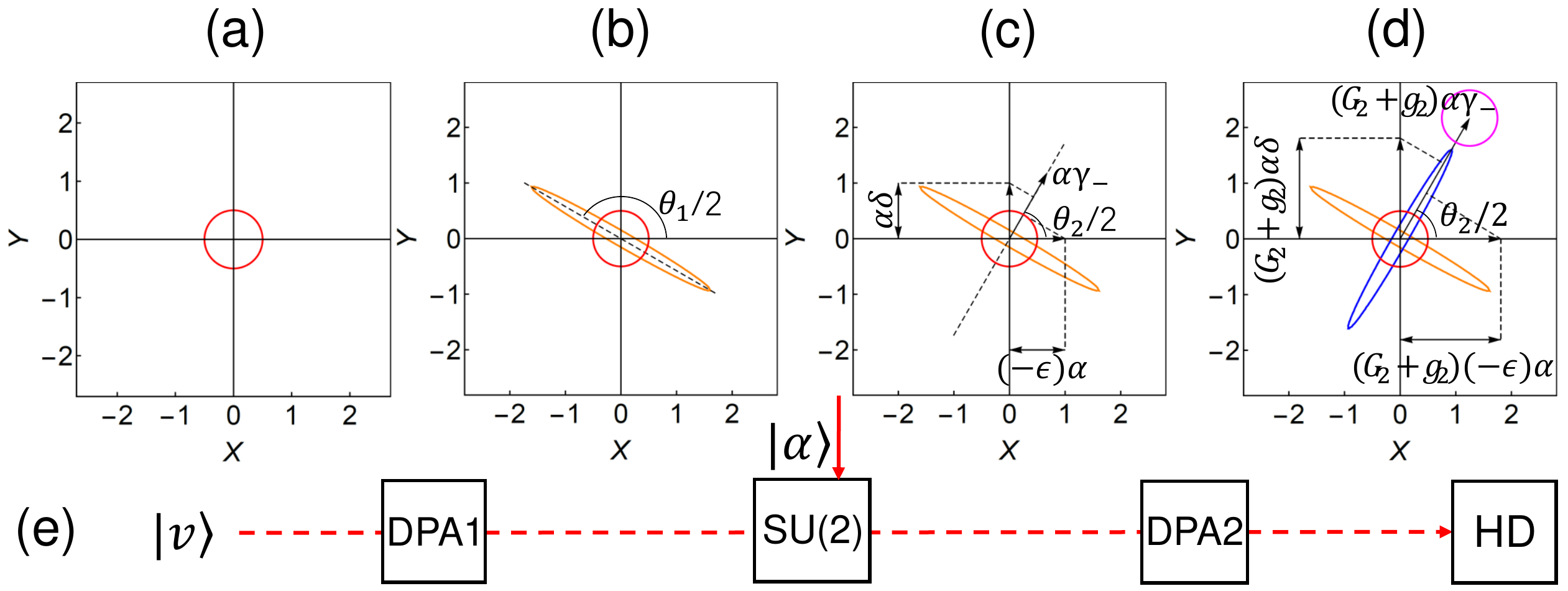}
	\caption{State evolution of the degenerate SU(1,1) interferometry. (a)-(d) are the phase-space (X-Y) representation of the evolved state at the corresponding position of the interferometer shown in (e), where $ \left| v \right\rangle  $  represents the vacuum state and $ \left| \alpha \right\rangle  $ represents the coherent state.}
	\label{fig:SUIstate}
\end{figure}

The physical meaning of the output signal in Eq.(\ref{eq:DPA2}) and noise in Eq.(\ref{N-DSUI}) for the degenerate SUI is illustrated in the phase space representation of the state evolution in Fig.\ref{fig:SUIstate} at each stage of the SUI: (a) shows the initial input vacuum state (circle); In (b) DPA1 squeezes the vacuum state and amplifies the vacuum noise at the selected direction ($\theta_1/2$) but de-amplifies in orthogonal direction; In (c) the SU(2) linear interferometer encodes weak signals  of phase $ \alpha\delta $ and amplitude $\alpha(-\epsilon) $ to the probe beam; In (d) DPA2 un-squeezes in the selected direction $\theta_2/2$ and noiselessly amplifies the mixed modulation signal $\alpha \gamma_-$. The actions of the two degenerate PAs are represented by two ellipses. Note that $\theta_{1,2}/2$ gives the direction of maximum amplification $(G_{1,2}+g_{1,2})^2$ of the phase-sensitive DPA1,2. The orthogonal orientations of the two ellipses is due to dark fringe operation condition: $(\theta_1-\theta_2)/2 = \pi/2$ for the SUI. When $G_1=G_2$, the noise in Eq.(\ref{N-DSUI}) becomes vacuum noise of one (circle) at the output because the actions of DPA1,2 are opposite but equal to each other (squeezing and then un-squeezing), leading to vacuum noise output, but in the same time DPA2 amplifies the modulation signal $\alpha\gamma_-$,  leading to sensitivity enhancement of $(G_1+g_1)^2$.

Different from non-degenerate SUI whose optimum sensitivity is achieved when $G_2\approx g_2\gg 1$, the degenerate SUI reaches optimum sensitivity at any gain $G_2$ for DPA2. On the other hand, to fully utilize the detection loss-tolerant property of SUI\cite{ou12,Marino,Manceau}, we need to have $G_2\gg G_1$ so that Eq.(\ref{N-DSUI}) gives output noise $\langle \Delta ^2\hat \cX \rangle = (G_2+g_2)^2/(G_1+g_1)^2 \gg 1$. In this way, any vacuum noise (size of 1) added through detection losses will be negligible compared to $\langle \Delta ^2\hat \cX \rangle$.

In summary, we have demonstrated that an SU(1,1) interferometer with a linear interferometer nested inside can make phase and amplitude measurement simultaneously with precision beating SQL, thus achieving quantum dense metrology. The phase and amplitude measurement share equally the overall resource of photon number and quantum entanglement. The degenerate version of the SUI can devote all resource to one measurement of an arbitrary mix of phase and amplitude with optimum measurement sensitivity. This should be particularly useful if the signal is encoded in both phase and amplitude such as LIGO interferometer where both phase and intensity fluctuations play important roles. In fact, we noticed a recent work from LIGO Collaboration\cite{Haocun} in which they experimentally demonstrated  quantum enhancement in a joint measurement of the phase of laser beams and position of mirrors by choosing proper squeezed angle of the initial squeezer (similar to $\theta_1$ in our work).

\noindent \textbf{Funding}\\
W.Z. acknowledge support by the National Key Research and Development Program of China under Grant number 2016YFA0302001, and the National Natural Science Foundation of China (NSFC) through Grant No. 11654005, Science and Technology Commission of Shanghai Municipality (STCSM) (16DZ2260200). J.F.C. acknowledge the support from NSFC through Grant No. 11674100 and Shanghai RisingStar Program No. 17QA1401300.

\noindent \textbf{Data availability statement}\\
Data sharing is not applicable to this article as no new data
were created or analyzed in this study.

\noindent \textbf{Disclosures}\\
The authors declare that there are no conflicts of interest related to this article.

\vskip 0.1in



%

\end{document}